# Soft Skills Requirements in Software Architecture's Job: An Exploratory Study

[1]Faheem Ahmed, [1]Piers Campbell, [1]Azam Beg, [2]Luiz Fernando Capretz

[1]Faculty of Information Technology, United Arab Emirates University

P. O. Box 17551, Al Ain, United Arab Emirates

[2]Department of Electrical & Computer Engineering, University of Western Ontario,
London, Ontario, Canada N6A 5B9

f.ahmed@uaeu.ac.ae, p.campbell@uaeu.ac.ae, abeg@uaeu.ac.ae, lcapretz@eng.uwo.ca

## Abstract

*The job of software architect is considered very crucial in the life cycle of the software development because software architecture has a pivotal role in the success and failure of the software project in terms of cost and quality. People have different personality traits, and the way they perceive, plan and execute any assigned task is influenced by it. These personality traits are characterized by soft skills. Most of the time, software development is a team work involving several people executing different tasks. The success and failure stories of software projects revealed the human factor as one of the critical importance. In this work we are presenting an exploratory study about the soft skills requirements for a software architect's job. We analyzed 124 advertised jobs in the title of software architect and explore the soft skills requirements. The results of this investigation help in understanding software skills requirement set for a job of software architect and presents the current status of their use in job advertisements.*

## 1. Introduction

Software architecture is gaining increasingly attention in the software development community due to its significantly important role in overall development cycle. Bass et al. [1] state that the software architecture of a program or computing system is the structure or structures of the system, which comprise software components, the externally visible properties of those components, and the relationships among them. A software architect as role participates individually or as a team member in creating software architecture. According to [2] responsibilities of an architect include articulating the architectural vision, conceptualizing and experimenting with alternative architectural approaches, creating models and component and interface specification documents, and validating the architecture against requirements and assumptions.

## 2. Research Motivations

Software engineering has been roughly characterized as set of activities comprising of phases such as system analysis, design, coding, testing, and maintenance. Logically they are different tasks which are coupled together to achieve the objective of software construction and operation. The micro-level interpretation of these activities demands a set of abilities of the people to carry them out effectively. For example, skills for designing a software system are quite different from those needed to test the software. The psychological hypothesis that not everyone can perform all tasks effectively reveals that personality traits play a critical role in the performance of people executing the same task.

Although there is huge diversity among software architecture principles, we can find common principles that are applicable to the architecture of any artifact, whether it is a poster, a household appliance, or a housing development. Although software architecture is still a young field and far from having a consensus on its relevant principles, software architecture will always requires the human creativity. Software architecture is an exploratory process, the architect looks for components by trying out a variety of schemes in order to discover the most natural and reasonable way to refine the solution. There has been a tendency to present software architecture in such a manner that it looks easy to do. Nevertheless, in the architecture of large and complex software, identification of key components is likely to take some time. Repetitions are not unusual, since a good architecture usually takes several iterations. The number of iterations also depends on the architect's insight and experience on the application domain.

Psychology asserts that not everybody is fit for all kind of tasks. We can achieve better results if we assign people with particular soft skills to different phases in a project. The advertisement of the jobs in the area of software engineering generally divides the skill requirements into two categories of "hard skills" and "soft skills". Hard skills are the technical requirements and knowledge a person should have to carry out a task. This includes the theoretical foundations and practical exposure a person should have to comfortably execute the planned task. Even though soft skills are the psychological phenomena which cover the personality types, social interaction abilities, communication, and personal habits, people believe that soft skills should complement the hard skills. The objective of this work is to review the job advertisements of software architect and explore the soft skills requirements. The results of this investigation help in understanding software skills requirement set for a job of software architect and presents the current status of their use in job advertisements.

## 3. Related Work

Many people believe that soft skills are essential in carrying out a job along with hard skills but software engineering community has not given much attention to this vital issue. A more concentration has been given to the hard skills requirements set. Few studies are carried out to look at the personality characteristics of software development life cycle title roles such as programmers, software engineers, and system analyst and have investigated the relationship between human skills and software life cycle phases. Using the 16PF test [3], Acuna et al. [4] measure the matching between individual capabilities (intrapersonal, organizational, interpersonal, and management) to software roles (team leader, quality manager, requirements engineer, designer, programmer, maintainer, tester, and configuration manager). Feldt *et al*. [5] evaluated the personality of 47 software professionals using the IPIP 50-item five-factor personality test [6]. After extensive statistical analyses they found that there are multiple and significant correlations between personality factors and software engineering, and concluded that individual differences in personality can explain and predict how judgments are made and decisions taken in software development projects. There is clear evidence that personality preferences have great impact on the motivation, quality of the work produced, and retention in the field of software engineering [7]. Bishop-Clark [8] investigated the relationship between cognitive aspects, personality traits and computer programming. Walz and Wynekoop [9] derive a methodology for identifying the traits and characteristics of top performing software developers.

## 4. Experimental Setup

In this exploratory study we visited some of the leading online job seeking website such as workopolis.ca, eurojobs.com and monster.ca. We looked at the jobs advertised under the title of "software architect". We collected a data of 124 job advertisements. We collected that data about nine soft skills includes communication skills, interpersonal skills, analytical and problem solving skills, team player, organizational skills, fast learner, ability to work independently, innovative and open and adaptable to changes. We found these nine soft skills most commonly used in the advertisement of the jobs. In order to better understand the usage and significance of these soft skills following paragraph provides some elaboration of these concepts.

Communication skills are the set of skills that enables a person to convey information so that it is received and understood. The term "interpersonal skills" is used often in business contexts to refer to the measure of a person's ability to operate within business organizations through social communication and interactions. Having positive interpersonal skills increases the productivity in the organization since the number of conflicts is reduced. Analytical and problem solving skill is the ability to visualize, articulate, and solve complex problems and concepts, and make decisions that make sense based on available information. Such skills include demonstration of the ability to apply logical thinking to gathering and analyzing information, designing and testing solutions to problems, and formulating plans. A team comprises a group of people linked in a common purpose and is especially appropriate for conducting tasks that are high in complexity and have many interdependent subtasks. A team player is a characteristic of an individual to work effectively in the team environment and contributes towards the desired goal. The ability to work independently narrates the individual's capability to carry out task with a minimal supervision. Innovation is process of being creativity by coming up with new ideas to resolve an issue in a much better way. Open and adaptable to changes reflects the personality of an individual to accept the changes in the carrying out tasks without showing resistance.

## 5. Data Processing & Results

We collected data from some well-known job advertising websites. The target population of the data consists of job titles such as software architect, senior software architect, system architect, lead software architect etc. We collected a data of 124 job advertisements. The jobs were generally belongs to various sectors ranges from wide range of operations such as consumer electronics, telecommunication, avionics, automobiles, financial, and information technology. The experience required for the jobs is classified into three categories of less than 3 years, 3to 6 years and greater than 6 years. Figure 1 illustrates the distribution of the data in terms of number of experience required for the job. 53% of the data belongs to jobs having required experience of less than 3 years. The jobs that required an experience of between 3 to 6 years constitute a data of about 31%. Whereas 16% of the dataset required an experience of more than 6 years.

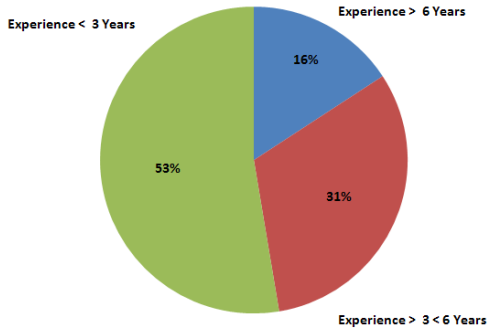

Figure 1: Data Distribution with respect to Experience Requirements

Table 1: Frequency Distribution of the Soft Skills Requirements

| Soft Skills | Yes | No |
|---|---|---|
| Communication Skills | 53% | 47% |
| Interpersonal Skills | 48% | 52% |
| Analytical & Problem Solving Skills | 55% | 45% |
| Organizational Skills | 34% | 66% |
| Fast Learner | 28% | 68% |
| Team Player | 51% | 49% |
| Ability to Work Independently | 30% | 70% |
| Innovative/Creative Mind | 46% | 54% |
| Open and Adoptable to Changes | 25% | 75% |

\* "Yes" = Job advertisement categorically required skill
"No" = Job advertisement did not categorically required skill

Table 1 illustrates the frequency distribution of the soft skills requirements advertisements in the job announcement. "Yes" column in the Table 1 indicates the percentage of the jobs where a specific soft skill was categorically required and "No" indicates that respective soft skill was not announced as requirement in the advertisement. The highest percentage of required skill was "analytical and problem solving" of about 55%. Lowest percentage of required skill was "Open and adoptable to changes" about 25%. The required soft skills of about less than 50% include "interpersonal", "organizational", "fast learner", "ability to work independently", "innovative and creative mind" and "open and adoptable to changes". "Communication skills", "analytical and problem solving" and "team player" skills are in high demand of requirements. Figure 2 illustrates the frequency distribution chart of the soft skills requirements in the job of software architect.

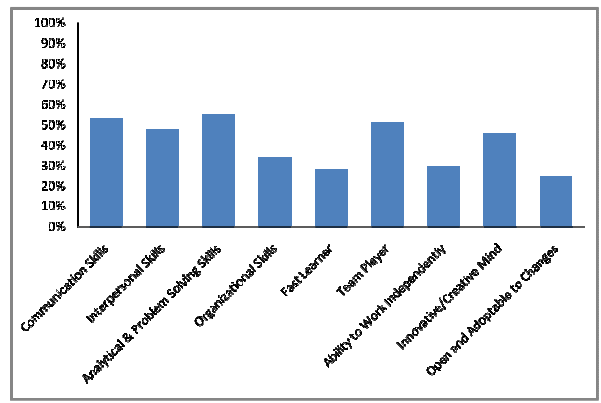

Figure 2: Frequency distributation of Soft Skills Requirements in Software Archtitect's Job

Table 2: Reliability of the Dataset

| Soft Skills | $R^2$ | Cronbach's Alpha |
|---|---|---|
| Communication Skills | 0.47 | 0.01 |
| Interpersonal Skills | 0.44 | 0.16 |
| Analytical & Problem Solving Skills | 0.49 | 0.42 |
| Organizational Skills | 0.23 | 0.28 |
| Fast Learner | 0.37 | 0.14 |
| Team Player | 0.20 | 0.20 |
| Ability to Work Independently | 0.22 | 0.22 |
| Innovative/Creative Mind | 0.21 | 0.19 |
| Open and Adoptable to Changes | 0.31 | 0.01 |

The two most important aspects of precision in survey based assessments are reliability and validity. Reliability refers to the reproducibility of a measurement, whereas validity refers to the agreement between the measured value and the true value. The potential threats to the external validity of survey-based assessments are the reliability and validity of the measuring instrument. The reliability of the survey was evaluated by using the internal-consistency analysis method. Internal-consistency analysis was performed using the coefficient alpha. We measured the construct validity using multiple squared correlations. Multiple squared correlations ($R^2$) are the percentage of response variable variation that is explained by its relationship with one or more predictor variables. In general, the higher the $R^2$, the better the model fits your data. Table 2 illustrates the reliability analysis of the survey. We observed positive multiple squared correlations where as cronbach alpha is observed from 0.01 to 0.49. The lower value of the cronbach's alpha indicates that they are disjoint variables.

## 6. Discussion

In the recent past software architecture has been gaining tremendous popularity in software development projects due to large scale project sizes. The objective of this work was to review the job advertisements of software architect and explore the soft skills requirements. The area identified by the study as in high demand is *Analytical and Problem Solving* skills. This skill is more in line with the job function of a software architect and as such one would expect to see this skill in high demand by the industry. The study identified that currently the industry is focused on the ability to communicate effectively and to function well as part of a team. In particular *communication skills* ranked highly as an in demand soft skill. Coupled with a demand of *interpersonal skills* we can clearly see that employers are placing a high value on the ability of employees to work well in groups internally to the organization but also externally with customers. Although the study identified *Organizational Skills* as being a soft skill in low demand by employers, but we assume that this skill points to an individual who is capable of managing their tasks, time and projects with potentially limited supervision. Coupled with the demand for high levels of communication and interpersonal skills, we can foresee that employers may require individuals who are capable of taking on a task as part of a team, articulating their work to colleagues and customers and operating with high levels of organization, but perhaps with limited supervision.

However the study identified a number of soft skills which are in surprisingly low demand by employers. The ability to be *a fast learner* is not in high demand. Neither is being a *ability to work independently* nor *open and adaptable to change*. These findings are surprising as these skills are related to the creative abilities, which are central to the creation of inventions such as software. However the study has shown that the role of software architect is significantly different from the typical software development roles. The emphasis in skills for an architect is in the areas of communication and management, rather than creativity and technical ability, which is relatively surprising.

Clearly the finding of the study highlight the differences of both role and function for software architects who are typically more closely involved with customers and senior management. This specialist role requires a dissimilar soft skill set to more typical software engineering roles, despite architects being involved in the same core function, the design of software products.

## 7. Conclusion

Software development is a team efforts. The team constitutes of humans therefore directly or indirectly the human factors which characterized the personality traits and soft skills are significantly important in team management. The objective of this work is to review the job advertisements of software architect and explore the soft skills requirements. The study found that some of the soft skills such as communication, team player, analytical and problem solving, and interpersonal are in demand even though not more that 50% whereas others are in lesser demand such as ability to work independently, fast learner and open and adaptable to changes. This study concludes that there is a requirement in the field of software engineering to understand the significance of soft skills requirements set.